\documentclass[
superscriptaddress,
article
 amsmath,amssymb,
 aps,
]{revtex4}
\usepackage{eucal}
\usepackage[utf8]{inputenc}
\usepackage{amsmath}
\usepackage[T1]{fontenc}
\usepackage{amsmath,amsfonts,amssymb,amsthm,epsfig,epstopdf,url,array}
\usepackage{soul}

\usepackage{dcolumn}% Align table columns on decimal point
\usepackage{bm}% bold math
\usepackage{graphicx}
\usepackage{comment}
\usepackage[caption=false]{subfig}

\newcommand{\bea}{\begin{eqnarray}}

\newcommand{\eea}{\end{eqnarray}}

\usepackage{color}

\begin{document}

\title{On the energy density in quantum mechanics}% Force line breaks with \\
\author{Francisco Torres Arvizu}
\affiliation{Instituto de Ciencias Físicas, UNAM, Av. Universidad s/n, CP 62210 Cuernavaca Morelos, México}
\author{Adrian Ortega}
\affiliation{Wigner RCP, Konkoly-Thege M. u. 29-33, H-1121 Budapest, Hungary}
\author{Hernán Larralde}
\address{Instituto de Ciencias Físicas, UNAM, Av. Universidad s/n, CP 62210 Cuernavaca Morelos, México}

%\date{\today}

\begin{abstract}

There are several definitions of energy density in quantum mechanics. These yield expressions that differ locally, but all satisfy a continuity equation and integrate to the value of the expected energy of the system under consideration. Thus, the question of whether there are physical grounds to choose one definition over another arises naturally. In this work, we propose a way to probe a system by varying the size of a well containing a quantum particle. We show that the mean work done by moving the wall is closely related to one of the definitions for energy density. Specifically, the appropriate energy density, evaluated at the wall corresponds to the force exerted by the particle locally, against which the work is done.  We show that this identification extends to two and three dimensional systems.

% In this paper we examine the ambiguity in the mathematical definition of energy density in
% quantum mechanics which, from a practical point of view, it contains a freedom of choice in its
% definition.[Could we soften this a little? What about:] We investigate the energy density in quantum
% mechanics which, from a practical point of view, it contains a freedom of choice in its definition.
% There are several expressions that differ locally that can be considered valid choices since they all
% integrate to the value of the expected energy of the system under consideration. Various authors
% have compared these different expressions in an attempt to establish whether there is one that is
% physically relevant or whether it comes down to a matter of preference. In this work, we propose a
% way to probe a system, by varying the size of an infinite well containing a quantum particle, and
% we show that the mean work done by moving the walls is closely related to one of the expressions
% for energy density. This highlights significant differences in the physical meaning at the local level
% between different candidates for the energy density. We show that our conclusions extend to two an
% three dimensional systems.

\end{abstract}

%\date{May 2022}

\maketitle

\section{Introduction}
In many textbooks on quantum mechanics \cite{schiff,merzbacher1961quantum, cohen1977quantum} we are taught that an important consequence that emerges from the fact that the wave function that describes a quantum particle satisfies the Schrodinger equation, is that the probability density for finding the particle at a given position satisfies a continuity equation.
This reflects the fact that probability is locally conserved \cite{merzbacher1961quantum, schiff,cohen1977quantum}. However, and rather surprisingly,  there is very little discussion in the literature regarding the concept of energy density in quantum mechanics, how to define it, how it evolves, whether it is conserved, and what is its physical significance \cite{bader,cohen79, cohen1984, Mathews, real_muga, muga_densities, MitaAJP2003, WuJPA2009, LudovicoPRB2014}.  In these few works, different expressions for the energy density have been proposed. These densities differ locally from one another, though they all integrate to the same value of the expected energy. It has been argued that there is no "physical" way to choose which of these different densities is physically meaningful, and that it essentially comes down to a matter of preference \cite{Mathews, real_muga}. In this paper we present what we believe are the minimum conditions that the energy density must satisfy and we discuss two different possible definitions that satisfy the conditions. We illustrate a way to probe the systems locally in order to identify which of the energy densities has a physical manifestation. To do this, we consider particles trapped in simple infinite well systems and probe the systems by changing the size of the well. We show that the expected work done to change the size of the well is directly related to the local value of one of the energy densities at the boundary. This density can be interpreted as the effective force exerted by the particle in the well. The other density is irrelevant in this respect as it vanishes identically at the walls. 

%\textcolor{blue}{[Just and idea:] This can be considered a quantum version of the classical radiation pressure of light~\cite{Jacksonbook1975}, albeit here the particle can have mass (e.g. it can be an electron but what is important is its wavelike properties).}

\section{The probability density and the energy density}
If we consider a quantum particle in a region $I$, in the position representation it is described by a wave function $\Psi(x,t)$ that satisfies the Schrodinger equation
\begin{eqnarray}
    i\hbar\frac{\partial}{\partial t} \Psi = \hat{H} \Psi \label{ShEq}
\end{eqnarray}
where
\begin{equation}
    \hat{H}=-\frac{\hbar^2}{2\mu} \nabla^2 + V(x)
\end{equation}
$V(x)$ is a potential and the wave function is normalized:
\begin{equation*}
    \int\limits_I |\Psi|^2 d\mathbf{x}=1.
\end{equation*}
The integrand $|\Psi(\mathbf{x},t)|^2\equiv\rho_D(\mathbf{x},t)$ is interpreted as the probability density of finding the particle at position $\mathbf{x}$ at time $t$. Then, as is well-known, the probability density fulfills a continuity equation

\begin{eqnarray}
    \frac{\partial}{\partial t} \rho_D +\nabla\cdot \mathbf{J_D}=0,
\end{eqnarray}
where the flux $\mathbf{J_D}$ is given by
\begin{equation}
      \mathbf{J_D}=\frac{\hbar }{2\mu i} (\psi ^* \nabla \psi-\psi\nabla \psi^*).
      \label{eqn:ced}
\end{equation}
This implies that the probability density is locally conserved \cite{schiff, cohen1977quantum}.
By the same token, the expected energy of a quantum particle can be expressed as
\begin{equation}
    \langle E\rangle=\int\limits_I \Psi^*\hat{H}\Psi ~d\mathbf{x} \label{Energy}
\end{equation}
This expression suggests that the integrand might be interpreted as an energy density, akin to the probability density discussed above.  If this was the case, the energy density in the system would be given by 
\begin{equation}
    \rho^E_1=\Psi^*\hat{H}\Psi=\Psi^*\left(-\frac{\hbar^2}{2\mu} \nabla^2 + V(x)\right) \Psi.
\end{equation}
Unfortunately, the choice $\rho^E_1$ above is not satisfactory as there is no guarantee that this density is real at every point where it is defined, though it is easy to see that the imaginary part always integrates to zero \cite{MitaAJP2003}.
% {In the standard case where $V(x)$ is real, the kinetic energy operator $\hat{p}^2/2m$ can be complex, since}
% \begin{equation}
%     \begin{split}
%     \Psi^* \frac{\hat{p}^2}{2m}\Psi &= -\frac{\hbar^2}{2m}\Psi^* \nabla^2 \Psi\\
%     &= -\frac{\hbar^2}{2m} \left(\nabla \cdot (\Psi^* \nabla\Psi) - \nabla \Psi^*\cdot \nabla \Psi \right),\\
%     \end{split} 
% \end{equation}
% \textcolor{blue}{where we have used the well-known identity $\Psi^.\nabla^2 \Psi = \nabla\cdot(\Psi^*\nabla \Psi) - \nabla\Psi^2\cdot\nabla\Psi$. The term proportional to $(\nabla \Psi^*\cdot \nabla \Psi)$ is positive definite; the term $\nabla \cdot (\Psi^* \nabla\Psi)$ is complex valued in general and integrates to zero over the volume of the system.}

It would be reasonable to expect that a physically acceptable definition of quantum energy density should, at least, meet the following conditions:
\begin{enumerate}
    \item it should be a local, real-valued function in the position representation;
    \item its integral over the system must be equal to the expected energy $\langle E\rangle$.
\end{enumerate} 
However, an infinite number of different expressions for the energy density satisfying these conditions can still be defined.  
%On the other hand, each expression corresponds to a density that is locally different from the rest. 

%Thus, to identify the a physically relevant expression, we need a way to measure a quantity related to the energy density locally in the system.   

%\textcolor{blue}{nor nonnegative??} 
% and a third possible requirement could be that
% \begin{enumerate}
%     \item[3.] It should be non-negative if the potential in the system is non-negative.
% \end{enumerate}

One possible definition of energy density, that does satisfy the above conditions, is
\begin{equation}
    \rho^E_2=\frac{1}{2}(\Psi^*\hat{H}\Psi + \Psi\hat{H}^*\Psi^*),
\end{equation}
which is obviously real, and integrates to $\langle E\rangle$. We cannot guarantee that $\rho_2^E$ is non-negative, even when the potential is positive for all $x$. Nevertheless, it has been argued that the negative kinetic energies that this entails are not necessarily unphysical \cite{Berry_2010}, and may even be measured \cite{Aharonov}.
Another possible choice for the energy density is %\textcolor{blue}{(see above)}
\begin{equation}
    \rho^E_3=\frac{\hbar^2}{2\mu} \left(\nabla\Psi^*\right)\cdot\left(\nabla \Psi\right) + \Psi^*V(x)\Psi
\end{equation}
which is real and also integrates to $\langle E\rangle$. In contrast to $\rho^E_2$, this choice is non negative if the potential $V(\mathbf{x})$ is non negative itself. 
%(many other densities satisfying the above conditions can be built as linear combinations of $\rho^E_2$ and $\rho^E_3$).

The energy densities defined above are locally conserved in the sense that they satisfy a continuity equation for appropriate energy fluxes:
\begin{eqnarray}
\begin{split}
   & \frac{\partial \rho_2^E}{\partial t}=-
    \frac{i\hbar}{4\mu }\nabla\cdot\Big(
    -\psi^*\nabla[H\psi]+\psi \nabla[H\psi^*]+(\nabla\psi^*)H\psi-(\nabla\psi)H\psi^*\Big)\\
\\
   & \frac{\partial\rho_3^E}{\partial t}
    =-\frac{i\hbar}{2\mu }\nabla\cdot\Big((\nabla\psi^*)H\psi-(\nabla\psi)H\psi^*  \Big)
    \end{split}
    \label{eqn:flux3}
\end{eqnarray}
Clearly, however, both expressions for the energy density differ locally, so the question of whether there are physical grounds to choose one over the other, or neither, arises naturally. Further, the physically meaningful density will be associated to an energy flux, which is central to the description of energy transport in quantum systems

In what follows, we propose a simple scheme to probe locally a particle in a box. Specifically, we calculate the work done by moving one of the box's walls a distance $d \ell $. The idea is that if this movement is done quickly, then the work will be due to interaction of the wall with the system at its immediate vicinity. As it turns out, the work done is independent of the speed at which the wall moves.
%, thus the motion can be thought of as being as fast as required so that the interaction is local.

\subsection{One dimensional system}

Following one of the procedures described in \cite{Doescher, Melnichuk_2005}, we begin by considering a particle of mass $\mu$ in a one dimensional infinite square well of size $L(t)$ (the extension to higher dimension "cubic" wells is straight forward). We can write the wave function of the particle as
\begin{equation}
    \Psi(x,t)=\sum\limits_{n=1}^\infty b_n(t) e^{-\frac{i}{\hbar}\int_0^t E_n(\tau)d\tau }\phi_n(x,t)
    \label{Psi}
\end{equation}
where $\phi_n(x,t)$ are the instantaneous eigenfunctions of the Hamiltonian,
\begin{equation}
    \phi_n(x,t)=\sqrt{\frac{2}{L(t)}} \sin{\left(\frac{n\pi x}{L(t)}\right)}\qquad n=1,2,...,
    \label{phi_n}
\end{equation}
corresponding to eigenvalues (instantaneous energies)
\begin{equation}
    E_n(t)=\frac{\hbar^2 n^2\pi^2}{2\mu L^2(t)},
\end{equation}
and the coefficients $b_n(t)$ are chosen so that $\Psi(x,t)$ satisfies the time dependent Schrodinger equation. Substituting $\Psi(x,t)$ into eq.(\ref{ShEq}) and using the ortonormality of the eigenfuctions $\phi_n(x,t)$, we find 
\begin{equation}
    \dot{b}_n(t)=-\sum_{n'=1}^{\infty} b_{n'}(t) e^{i({\theta_n(t)-\theta_{n'}(t)})} \int_{0}^{L(t)} \phi_n(x,t) \dot{\phi}_{n'}(x,t) dx \label{bdot}
\end{equation}
where $\theta_n(t)=\frac{1}{\hbar}\int_0^t E_n(\tau)d\tau$. The initial values of the $b_n(t)$ are fixed when we prepare the initial state:
\begin{equation}
    \Psi(x,0)=\sum\limits_{n=1}^\infty b_n(0)\phi_n(x,0)
    \label{Psi0}.
\end{equation}
The expected energy of the particle is given by
\begin{equation*}
\langle\Psi|\hat{H}|\Psi\rangle\equiv\mathcal{E}(t)=\sum_{n=1}^\infty|b_n(t)|^2E_n(t).\label{expectedE}
\end{equation*}
We are interested in the change in energy of the particle due to the motion of the wall after a time $\delta t$, i.e. $\mathcal{E}(\delta t)-\mathcal{E}(0)\approx\dot{\mathcal{E}}(0)\delta t$. From eq.(\ref{expectedE}) we can calculate
\begin{equation}
 \dot{\mathcal{E}}(0)=\sum_{n=1}^\infty E_n(0) (b^*_n(0)\dot{b}_n(0)+\dot{b}^*_n(0)b_n(0)) +\dot{E_n}(0)|b_n(0)|^2
\end{equation}
where $\dot{E}_n(0)$ is 
\begin{equation}
    \dot{E}_n(0)=-\frac{\hbar^2n^2\pi^2}{\mu L(0)^3}v,
\end{equation}
$L(0)$ is the initial length of the well, and $v\equiv \dot{L}(0)$ is the initial speed at which the wall moves. The $\dot{b}_n(0)$ can be calculated using eq. (\ref{bdot}), where the time derivatives of the instantaneous eigenfunctions can be obtained from equation (\ref{phi_n}) and the resulting integrals appearing in eq. (\ref{bdot}) can be evaluated directly

\begin{equation}
   \int_{0}^{L(t)} \phi_{n'}(x,t) \dot{\phi_{n}}(x,t) dx= \begin{cases}  \frac{2nn' (-1)^{n-n'} }{L(t)(n'^2 -n^2)}  v,  &n'\neq n,\\ \\
   0, & n'=n.
   \end{cases}
\end{equation}
Then, after a little algebra, $\dot{\mathcal{E}}(0)$ can be expressed in terms of the initial conditions of the system as
% \begin{equation}
% \begin{split}
%   \sum_{n} E_n(0) (b^*_n(0)\dot{b}_n(0)+\dot{b}^*_n(0)b_n(0)) &=-\sum_{n}\sum_{n'\neq n} b_{n}(0)b^*_{n'}(0) \frac{\hbar^2 \pi^2}{2mL_0^2} (n'^2-n^2) \left[ \frac{2nn' (-1)^{n-n'} }{L_0(n'^2 -n^2)} \right]\\
%   &= -\frac{\hbar^2 \pi^2}{m L_0^3 } v\sum_{n}\sum_{n'\neq n} b_{n}(0)b^*_{n'}(0) (-1)^{n-n'} nn´'
% \end{split} 
% \end{equation}
% in the last expression $n\neq n'$. Susutuing the above expressions we have that  $\dot{\mathcal{E}}(0)$ is
% \begin{equation}
%     \dot{\mathcal{E}}(0)= -\frac{\hbar^2 \pi^2}{\m L_0^3 } v
%     \left\{ \sum_{n}|b_n(0)|^2 n^2 +\sum_{n}\sum_{n'\neq n} b_{n}(0)b^*_{n'}(0) (-1)^{n-n'} nn´'\right\} 
% \end{equation}

\begin{equation}
    \dot{\mathcal{E}}(0)= -\frac{\hbar^2 \pi^2 v}{\mu L(0)^3 } 
    \left| \sum_{n=1}^{\infty} b_{n}(0)(-1)^{n} n\right|^2
\end{equation}

Now we compare this expression with the value of the value of energy density $\rho_2^E$ and $\rho_3^E$ evaluated at the wall. For this system we have
\begin{equation}
\begin{split}
\rho_2 ^E(x,t)&= \frac{\hbar^2 \pi^2 }{2\mu L(t)^3} \mathbf{\mathrm{Re}} \left[ \sum_{n=1}^{\infty} \sum_{ n'=1}^{\infty} n^2 b^*_{n'}(t) b_{n}(t) e^{i(\theta_{n'}(t)-\theta_n(t))}\sin\left(\frac{ n'\pi x}{L(t)}\right) \sin\left(\frac{n \pi x}{L(t)} \right)\right].
\end{split}
\end{equation}
Evaluating at the wall, $x=L(t)$, we see that $\rho_2 ^E(L(t),t)=0$ trivially for all time. On the other hand, for $\rho_3(x,t)$ we have
\begin{equation}
 \rho_3^E(x,t)=\left(\frac{\hbar^2 }{\mu L(t)}\right)\left[\sum_{n'=1}^{\infty} e^{i\theta_{n'}(t)}b_{n'}^*(t) \frac{n'\pi }{L(t)}\cos\left(\frac{n' \pi x}{L(t)}\right) \right ]   \left[\sum_{n=1}^{\infty} e^{-i\theta_n(t)}b_n(t) \frac{n\pi }{L(t)}\cos\left(\frac{n \pi x}{L(t)} \right)\right ].
\end{equation}
Evaluating this expression at the wall, $x=L(t)$, at $t=0$ we get
\begin{equation}
 \rho_3^E(L(0),t=0)   = \frac{\hbar^2 \pi^2}{\mu L(0)^3 } 
    \left| \sum_{n=1}^{\infty} b_{n}(0)(-1)^{n} n\right|^2.
\end{equation}
Thus, the average amount of work $\delta \mathcal{W}$ done by the the motion of the wall during a time interval $\delta t$ is 
\begin{equation}
    \delta \mathcal{W}=\mathcal{E}(\delta t)-\mathcal{E}(0)=\dot{\mathcal{E}} (0) \delta t=-\rho_3^E(L(0),0)   \delta \ell
    \label{eq:dW}
\end{equation}
where $\delta \ell=v\delta t$ is the distance the wall moves. Both the particle density $\rho_D$ and $\rho_2^E$ vanish at the wall, thus, this result suggests that the wall transfers energy to, or receives energy from the particle as if the wall interacted with the system through the energy density $\rho_3^E$. In particular, in this one dimensional example, $\rho_3^E(L(0),0)$, being an energy per unit length, can be thought of as the mean force exerted locally and instantaneously by the particle on the wall. 

\subsection{Two and three dimensional systems}

For completeness, we show that these results extend to a quantum particle in a circular box in two dimensions and in a spherical box in three dimensions.

In the two dimensional case we consider a particle with mass $\mu $ confined in an infinite circular potential
\begin{equation}
    \begin{array}{cc}
 V(r)= &  \left\{
    \begin{array}{cc}
      0,  \quad\text{if}& r\leq  R(t), \\
      \infty,  \quad\text{if}& r>R(t),
    \end{array}
    \right.
\end{array}
\end{equation}
where the radius varies in time $R(t)=R_0+vt$. The wave function takes the form 
\begin{equation}
    \Psi(\vec{r},t)= \sum_{m=-\infty}^{\infty}\sum_{n=1}^{\infty} b_{m,n}(t) e^{-i\theta_{m,n} (t)} \frac{1}{\sqrt{\pi} R(t) J_{m+1}(z_{m,n})}J_m\left(k_{m,n} r \right) e^{im\phi} ,
\end{equation}
where $z_{m,n}$ is the nth the zero of $J_m(x)$, the Bessel function of order $m$, $ k_{m,n}=\frac{z_{m,n}}{R(t)}$ and  $\theta_{m,n} (t)= \frac{1}{\hbar} \int_{0}^{t} E_{m,n} (t')dt'$ as before, with 
\begin{equation}
    E_{m,n}(t)= \frac{\hbar ^2 z_{m,n}^2}{ 2\mu R(t)^2}.
\end{equation}
Following the same procedure as above (see details in Appendix A), we find that the average work done to move the wall a distance $\delta \ell= v\delta t$ is given by 
\begin{equation}
      \dot{\mathcal{E}}(0) \delta t =-\frac{\hbar^2}{\mu  R_0^3}  \delta \ell \sum_{m=-\infty}^{\infty}\left|\sum_{n=1}^{\infty}b_{m,n}(0)z_{m,n}\right|^2 . 
\end{equation}
Now we check that this work is done against the mean total force exerted by the particle on the wall through the energy density $\rho_3^E$ :
\bea
\mathcal{F}\equiv \int \rho_3^E(\vec{r},t=0) \delta(r-R_0)r drd\phi \label{int2d}
\eea
where
\begin{equation}
\begin{split}
\rho^E_3(\vec{r},t)&= \frac{h^2}{2\mu}\sum_{m=-\infty}^{\infty}\sum_{n=1}^{\infty} \sum_{m'=-\infty}^{\infty}\sum_{n'=1}^{\infty}  e^{(m-m')i \phi}\frac{ b_{m,n}(t)b_{m',n'}^*(t) }{\pi R(t)^2 J_{m+1 } \left(z_{m,n}\right)J_{m'+1}  \left(z_{m',n'}\right)} e^{i({\theta_{m,n'}(t)-\theta_{m,n}(t)})} \\&\times\left[\frac{\partial}{\partial r}  J_m  \left(k_{m,n}r\right) \frac{\partial}{\partial r} J_{m'}  \left(k_{m',n'}r\right) +   \frac{mm'}{r^2} J_m  \left(k_{m,n}r\right) J_{m'}\left(k_{m',n'}r\right)  \right]
\end{split}
\end{equation}
Evaluating at $t=0$ and performing the integral in eq. (\ref{int2d}), we find
\begin{equation}
   \mathcal{F}= \frac{h^2}{\mu R_0^3 } \sum_{m=\infty}^{\infty}  \left|\sum_{n=1}^{\infty}  b_{m,n}(0) z_{m,n}\right|^2. 
\end{equation}
so indeed we have $\delta\mathcal{W}=-\mathcal{F}\delta\ell$

Finally, for the three dimensional case we consider a particle  with mass $\mu $ in an infinite spherical moving well potential of radius $R(t)=R_0+vt$. 
Now the wave function takes the form 
\begin{equation}
    \Psi(\vec{r},t)= \sum_{n=1}^{\infty}\sum_{l=0}^{\infty} \sum_{m=-l}^{l}b_{n,l,m}(t) e^{-i\theta_{l,n} (t)} \frac{\sqrt{2}}{R(t)^{3/2} j_{l+1}(z_{l,n})}j_l\left(k_{l,n} r \right) Y_{l,m}(\theta, \phi),
\end{equation}
where $z_{l,n}$ is the nth zero of $j_l(x)$, the spherical Bessel function of order $l$, $Y_{l,m}(\theta, \phi) $ are the spherical harmonics, and $\theta_{l,n} (t)= \frac{1}{\hbar} \int_{0}^{t} E_{l,n} (t')dt'$, as before. For this case, the change in the total energy of the system caused by moving the boundary at a speed $v$ for a time $\delta t$ is (See Appendix B):
\begin{equation}
   \dot{\mathcal{E}}(0)\delta t= -\frac{h^2 v \delta t}{\mu R_0^3 } \sum_{n'=1}^{\infty}\sum_{n=1}^{\infty}\sum_{l=0}^{\infty}\sum_{m=-l}^{m=l}    b_{n,l,m}(0)b_{n',l,m}^*(0) z_{l,n}z_{l,n'}
\end{equation}

% \textcolor{red}{One of my questions still is if the energy density is an observable. Eq. (20) shows that essentially the work done by the motion of the wall is proportional to $\rho_3^E$, but here one knows that "Fluctuation theorems: Work is not an observable" [PHYSICAL REVIEW E 75, 050102 (2007)]. If the energy density is not an observable, what do we seek to describe?}

As in the previous examples, this change of energy is due to work done against the total force exerted by the particle on the boundary. In this case this is given by 
\bea
\mathcal{F}=\int\rho_3(\vec{r},t=0) \delta(r-R_0)r^2 drd\Omega . 
\eea
where
\begin{equation}
\begin{split}
\rho_3^E(\vec{r},t)&= \frac{h^2}{2\mu }\sum_{n=1}^{\infty}\sum_{l=0}^{\infty}\sum_{m=-l}^{m=l}\sum_{n'=1}^{\infty}\sum_{l'=0}^{\infty}\sum_{m'=-l'}^{m'=l'}  \frac{2 b_{n,l,m}(t)b_{n',l',m'}^*(t) }{R(t)^3  j_{l'+1 } \left(z_{l',n'}\right)j_{l+1}  \left(z_{l,n}\right)} e^{i({\theta_{l,n'}(t)-\theta_{l,n}(t)})} \\&\times\left[\frac{\partial}{\partial r}  j_l  \left(k_{l,n}r\right) \frac{\partial}{\partial r} j_{l'}  \left(k_{l',n'}r\right) \mathbf{Y}_{l',m'}^* \cdot \mathbf{Y}_{l,m}+   \frac{1}{r^2} j_l  \left(k_{l,n}r\right) j_{l'}\left(k_{l',n'}r\right) \mathbf{\Psi}_{l,m}^*\cdot \mathbf{\Psi}_{l',m'} \right].
\end{split}
\end{equation}
Here $ \mathbf{Y}_{l,m}= Y_{l,m}(\theta,\psi)\hat{r} $ and $\mathbf{\Psi}_{l,m}=r \nabla Y_{l,m}(\theta,\psi)$ are the vector spherical harmonics~\cite{RBarrera_1985}. These satisfy the orthogonality relation $\mathbf{\Psi}_{l,m}\cdot\mathbf{Y}_{l,m} $ =0, and the integral relations $\int_{0}^{2\pi}\int_{0}^{\pi} \mathbf{\Psi}_{l,m}^*\cdot \mathbf{\Psi}_{l',m'} d\Omega=l(l+1) \delta_{l,l'}\delta_{m,m'}$ and $\int_{0}^{2\pi}\int_{0}^{\pi} \mathbf{Y}_{l,m}^*\cdot \mathbf{Y}_{l',m'}d\Omega=\delta_{l,l'}\delta_{m,m'}$. Thus, evaluating $\rho_3(\vec{r},t)$ at $t=0$ and integrating, we have 
\begin{equation}
  \mathcal{F}=\frac{h^2}{2\mu } \sum_{n'=1}^{\infty}\sum_{n=1}^{\infty}\sum_{l=0}^{\infty}\sum_{m=-l}^{m=l}   \frac{2 b_{n,l,m}(0)b_{n',l,m}^*(0) z_{l,n}z_{l,n'}}{R_0^3  j_{l+1 } \left(z_{l,n}\right)j_{l+1}  \left(z_{l,n'}\right)}\left[ j'_l  \left(z_{l,n}\right)  j'_l  \left(z_{l,n'}\right) \right]. 
\end{equation}
Finally, using the properties of the Bessel functions, we get
\begin{equation}
   \mathcal{F}= \frac{h^2}{\mu R_0^3 } \sum_{l=0}^{\infty}\sum_{m=-l}^{m=l}  \sum_{n'=1}^{\infty}\sum_{n=1}^{\infty}  b_{n,l,m}(0)b_{n',l,m}^*(0) z_{l,n}z_{l,n'}
= \frac{h^2}{\mu R_0^3 } \sum_{l=0}^{\infty}\sum_{m=-l}^{m=l} \left|\sum_{n=1}^{\infty}  b_{n,l,m}(0) z_{l,n}\right|^2
\end{equation}
Thus, we again have $d\mathcal{W}=-\mathcal{F}d\ell$, or equivalently,
$d\mathcal{W}=-\mathcal{P}dV$, where the pressure is $\mathcal{P}=\mathcal{F}/A$, the differential of volume is $dV=A d\ell$, and $A=4\pi R_0^2$ is the area of the surface.
\section{Discussion}

If we have a quantum particle in a infinite well, changing the size of the well changes the energy of the system. Conservation of energy implies that this change of energy must be due to work done when moving the walls against the force exerted by particle. We have shown that the mean amount of work done when moving the wall is directly related to the energy density $\rho_3^E$ at the wall. This energy density also satisfies the physical requirements of being local, and, of course, integrates to the expected value of energy. Further, this expression yields a positive definite value everywhere if the potential is also positive, and it satisfies a continuity equation, implying that it is locally conserved. In contrast, due to the boundary condition $\rho_2^E$ vanishes identically at the wall (as does the probability density $\rho_D$), so its value cannot be probed by moving the wall, and though it is also conserved, it is not necessarily positive everywhere it is defined, even when the potential is nowhere negative, which implies that the particle may have negative kinetic energy in some regions of the system.
We contend that these results imply that $\rho_3^E$ can be considered as the actual physical energy density of the system, in the sense that it is measurable as the mean instantaneous local force (pressure) exerted by the particle on the wall. This would also imply that the energy flux associated to $\rho_3^E$ is the relevant flux to describe energy transport in quantum systems.
From a different perspective, $\rho^E_3$ is analogous to the electromagnetic radiation pressure on a perfect absorber, as in that case too $P_{\rm rad}= u$, where $u$ is the energy density of the wave. Of course, if all that is wanted is the integral of the density to obtain the expected energy, then any one of the expressions discussed in this work can be used, even $\rho_1^E$, which may have complex values locally.

\section*{Acknowledgements}

Francisco Torres Arvizu acknowledges support from CONAHCYT scholarship number 834573. A. Ortega acknowledges support by the Ministry of Culture and Innovation and the National Research, Development and Innovation Office within the Quantum Information National Laboratory of Hungary (Grant No. 2022-2.1.1-NL-2022-00004).

% \textcolor{blue}{There are two things that we might include. Both of these ideas I draw them from here \url{https://chem.libretexts.org/Bookshelves/Physical_and_Theoretical_Chemistry_Textbook_Maps/Quantum_Tutorials_(Rioux)/01\%3A_Quantum_Fundamentals/1.98\%3A_Quantum_Mechanical_Pressure}}
% \begin{itemize}
%     \item Usually, the pressure is defined as $P = -d K/dV$, where $K$ is the kinetic energy and $V$ is the volume of the system. In our case, this would be equivalent to $$P = -\frac{dK}{dL} = \frac{\hbar^2 n^2 \pi^2}{\mu L^3}.$$ Is it not that Eq. (22) here must be this, i.e. $\rho_3^E = \frac{\hbar^2 n^2 \pi^2}{\mu L^3}$?

%     \item Something that the webpage mentions is that, rather obviusly, reducing the size of the well increases "dramatically" the value of the kinetic energy. Further "This “repulsive” character of quantum mechanical kinetic energy is the ultimate basis for the stability of matter". Well, I've never saw such argument and seems rather nice, perhaps we can discuss it further.

%     %%\item In the 3d case, is it not that, depending on the initial energy of the particle, the pressure exerted is not always uniform? If I think that the eigenfunction of the particle is defined \textbf{only} with spherical harmonics, then the probability distributions in general are not spherical symmetric. This would mean that the particle "touches the walls" with probability approaching to zero (as in the one dimensional box), while in other cases it will simply never touch the wall. Hence, the pressure exerted might depend in the eigenfunction the particle posses., right?
%     %Perhaps we can make a plot about this.
% \end{itemize}

\appendix
\section{Two Dimensional case}
Consider a particle  with mass $\mu $ in an infinite spherical potential moving well 
\begin{equation}
    \begin{array}{cc}
 V(r)= &  \left\{
    \begin{array}{cc}
      0  \quad\text{if}& r\leq  R(t) \\
      \infty  \quad\text{if}& r>R(t)
    \end{array}
    \right.
\end{array}
\end{equation}
where $R(t)=R_0+vt$. The Hamiltonian operator is 
\begin{equation}
    \mathcal{H}=-\frac{\hbar^2}{2\mu } \left[\frac{1}{r} \frac{\partial }{\partial r} r \frac{\partial }{\partial r} +\frac{1}{r^2} \frac{\partial ^2}{\partial \phi^2}\right], 
\end{equation}
in this case, the instantaneous eigenfunctions are 
\begin{equation}
    \psi_{n,l,m}(r,\phi,t)=\frac{1}{\sqrt{\pi} R(t) J_{m+1}(z_{m,n})}J_m\left(k_{m,n} r \right) e^{im\phi},  
 \qquad k_{m,n}=\frac{z_{m,n}}{R(t)}, 
\end{equation}
where $z_{m,n}$ is the nth the zero of $J_m(x)$, the Bessel function of order $m$. The wave function takes the form 

\begin{equation}
    \Psi(\vec{r},t)= \sum_{m=-\infty}^{\infty}\sum_{n=1}^{\infty} b_{m,n}(t) e^{-i\theta_{m,n} (t)} \frac{1}{\sqrt{\pi} R(t) J_{m+1}(z_{m,n})}J_m\left(k_{m,n} r \right) e^{im\phi} ,
\end{equation}
where $\theta_{m,n} (t)= \frac{1}{\hbar} \int_{0}^{t} E_{m,n} (t')dt'$, with 
\begin{equation}
    E_{m,n}(t)= \frac{\hbar ^2 z_{m,n}^2}{ 2\mu R(t)^2}.
\end{equation}
Then, the expected energy is given by 
\begin{equation*}
\langle\Psi|\hat{H}|\Psi\rangle\equiv\mathcal{E}(t)=\sum_{m=-\infty}^{\infty}\sum_{n=1}^{\infty} |b_{m,n}(t)|^2E_{m,n}(t) , 
\end{equation*}
and its derivative at $t=0$
\begin{equation}
 \dot{\mathcal{E}}(0)=\sum_{m=-\infty}^{\infty}\sum_{n=1}^{\infty} \left\{E_{m,n}(0) \Big(b^*_{m,n}(0)\dot{b}_{m,n}(0)+\dot{b}^*_{m,n}(0)b_{m,n}(0)\Big) +\dot{E}_{m,n}(0)|b_{m,n}(0)|^2 \right\},
\end{equation}
where $\dot{b}(t)$ and $\dot{E}_{m,n}(t)$ are 
\begin{equation}
    \dot{E}_{m,n}=-v\frac{\hbar ^2 z_{m,n}^2}{ \mu R(t)^3}, 
\end{equation}
\begin{equation}
    \dot{b}_{m,n}(t)=-\sum_{m'=-\infty}^{\infty}\sum_{n'=1}^{\infty}  b_{m',n' }(t) e^{i({\theta_{m,n}(t)-\theta_{m',n'}(t)})} \int_{0}^{R(t)} \int_{0}^{2\pi}  \psi_{m,n}(r,\phi,t)\dot{\psi}_{m',n'}(r,\phi,t) rdrd\phi \label{bdot2a},
\end{equation}
respectively. The integral in the last expression when $n=n'$ satisfies
\begin{equation}
   \int_{S_0} \psi_{n}\frac{\partial }{\partial t} \psi^*_{n} dS=-\int_{S_0} \psi_{n}^*\frac{\partial }{\partial t} \psi_n dS
\end{equation}
which implies that these terms in $\dot{\mathcal{E}}(0)$ vanish, moreover when $n\neq n'$ we have
\begin{equation}
   \int_{0}^{R(t)}\int_{0}^{2 \pi }  \psi_{m,n}(r,\phi,t)\dot{\psi}_{m',n'}(r,\psi,t) rdrd\phi = \delta_{m,m'}  \frac{2}{R(t)^2 J_{m+1}(z_{m,n}) J_{m'+1}(z_{m',n'})}\int_{0}^{R(t)} J_m\left(k_{m,n} r \right) \dot{J}_{m'}\left( k_{m',n'} r\right) rdr, 
\end{equation}
substituting
\begin{equation}
\begin{split}
\sum_{m=-\infty}^{\infty}\sum_{n=1}^{\infty}  \ E_{m,n}(t) \dot{b}_{m,n}(t)b^*_{m,n}(t)&=  - \sum_{m=-\infty}^{\infty}\sum_{ n'=1}^{\infty}\sum_{ n=1}^{\infty}b_{m,n'}(t)b^*_{m,n}(t)E_{m,n}(t)  e^{i({\theta_{m,n}(t)-\theta_{m,n'}(t)})} \times\\ & \qquad\frac{2}{R(t)^2 J_{m+1}(z_{m,n}) j_{m'+1}(z_{m,n'})} \int_{0}^{R(t)} J_l\left(k_{m,n} r \right) \dot{J}_{l}\left( k_{m,n'} r\right) rdr, 
\\ & =-  \sum_{m=-\infty}^{\infty}  \sum_{n=1}^{\infty}\sum_{n'=1}^{\infty}b^*_{m,n'}(t)b_{m,n}(t) E_{m, n'}(t)  e^{i({\theta_{m,n'}(t)-\theta_{m,n}(t)})} \times \\ & \qquad \frac{2}{R(t)^2 J_{m+1}(z_{m,n}) J_{m'+1}(z_{m,n'})} \int_{0}^{R(t)} J_m\left(k_{m,n'} r \right) \dot{J}_{m}\left( k_{m,n} r\right) rdr,
\end{split}
\end{equation}
and
\begin{equation}
\begin{split}
 \sum_{m=-\infty}^{\infty}\sum_{n=1}^{\infty}  E_{m,n}(t) \dot{b}^*_{m,n}(t)b_{m,n}(t)& = -\sum_{m=-\infty}^{\infty} \sum_{n=1}^{\infty}\sum_{n'=1}^{\infty}  b_{m,n}(t)b^*_{m,n'}(t)E_{m,n}(t)  e^{i({\theta_{m,n'}(t)-\theta_{m,n}(t)})} \times \\ & \qquad \frac{2}{R(t)^2 J_{m+1}(z_{m,n}) J_{m+1}(z_{m,n})} \int_{0}^{R(t)} J_m\left(k_{m,n} r \right) \dot{J}_{n}\left( k_{m,n'} r\right) rdr.
\end{split}
\end{equation}
Thus, 
\begin{equation}
\footnotesize
   \frac{2}{R(t)^2 J_{m+1}(z_{m,n}) J_{m+1}(z_{m,m'})} \int_{0}^{R(t)} J_l\left(k_{m,n} r\right) \dot{J}_m\left(k_{m,n'} r\right) rdr=  -\frac{2 z_{m,n'}  v}{J_{m+1} \left(z_{m,n}\right)  J_{m+1} \left(z_{m,n'}\right)R(t)^4}\int_{0}^{R(t)} J_{m} \left(k_{m,n} r \right)  J'_{m} \left
    (k_{m,n'} r\right) r^2 dr,
\end{equation}
 using \footnote{ Using Abramowitz 11.3.9 
$$ \int_{z} \Big\{  (k^2-l^2) t - \frac{(\mu^2-\nu^2)}{t}\Big \}\mathcal{C} _{\mu} (k t)\mathcal{D}_{\nu}(l t )dt =z\{ k \mathcal{C}_{\mu+1} (kz)\mathcal{D}_{\nu}(lz)-l\mathcal{C}_{\mu} (kz)\mathcal{D}_{\nu+1}(lz)\} -(\mu-\nu) \mathcal{C} _{\mu} (k t)\mathcal{D}_{\nu}(l t )$$ 
for two cylindrical functions $\mathcal{C}_{\mu} (k t)$ and $\mathcal{D}_{\nu} (k t) $, \cite{abramowitz1965handbook}} and the Leibnitz rule, we obtain
 \begin{equation}
     \int_{0}^{R(t)} J_{m+1} \left(k_{m,n} r\right)  J'_{m+1} \left(k_{m,n' } r\right) r^2dr= -\frac{R^3(t) z_{m,n}}{z_{m,n'}^2-z_{m,n}^2} J_{m+1} \left(z_{m,n}\right)  J'_{m} \left(z_{m,n'}\right). 
     \label{eqn:j}
 \end{equation}
By the recurrence relation $J'_p(x)=\frac{p}{x}J_p(x)-J_{p+1}(x)$, 
\begin{equation}
  \int_{0}^{R(t)} J_{m} \left(k_{m,n} r\right) J_{m}'  \left(k_{m,n' } r\right) r^2dr= \frac{R^3(t) z_{m,n}}{z_{m,n'}^2-z_{m,n}^2}  J'_{m} \left(z_{m,n}\right)  J'_{m} \left(z_{m,n'}\right),
 \end{equation}
analogously we have
\begin{equation}
  \int_{0}^{R(t)} J_{m} \left(k_{m,n'} r\right) J_{m}'  \left(k_{m,n } r\right) r^2dr=-\frac{R^3(t) z_{m,n}}{z_{m,n}^2-z_{m,n'}^2}  J'_{m} \left(z_{m,n}\right)  J'_{m} \left(z_{m,n'}\right),
 \end{equation}
 evaluating $\dot{\mathcal{E}(t)}$ at $t=0$
\begin{equation}
      \dot{\mathcal{E}}(0)=-\frac{\hbar^2}{\mu  R_0^3} v \left\{\sum_{m=-\infty}^{\infty}\sum_{n=1}^{\infty}z_{m,n}^2|b_{m,n}(0)|^2+ \sum_{m=-\infty}^{\infty}\sum_{n=1}^{\infty}\sum_{n'\neq n}^{\infty} b^*_{m,n}(0)b_{m,n'}(0) z_{m,n}z_{m, n'} \frac{J'_{m} \left(z_{m,n}\right)  J'_{m} \left(z_{m,n'}\right)}{J_{m+1} \left(z_{m,n}\right)  J_{m+1} \left(z_{m,n'}\right)} \right\}, 
\end{equation}
using the recurrence formula again
\begin{equation}
      \dot{\mathcal{E}}(0)=-\frac{\hbar^2}{\mu  R_0^3} v \left\{\sum_{m=-\infty}^{\infty}\sum_{n=1}^{\infty}z_{m,n}^2|b_{m,n}(0)|^2+ \sum_{m=-\infty}^{\infty} \sum_{n=1}^{\infty}\sum_{n'\neq n}^{\infty} b^*_{m,n}(0)b_{m,n'}(0) z_{m,n}z_{m,n'} \right\}. 
\end{equation}
On the other hand, the energy density is given by 
\begin{equation}
\rho_3(\vec{r},t)= \frac{\hbar^2}{2\mu} \nabla \Psi^*(\vec{r},t)\cdot \nabla \Psi(\vec{r},t) 
\end{equation}
in this case we take 
\begin{equation}
    \nabla \Psi(\vec{r},t)=\sum_{m=-\infty}^{\infty}\sum_{n=1}^{\infty} e^{-i\theta_{m,n}} b_{m,n}(t) \frac{e^{im\phi}}{\sqrt{\pi}R(t) J_{m+1}(z_{m,n})}  \left[\frac{\partial}{\partial r}  J_m  \left(z_{m,n}r\right) \hat{r}  +   \frac{i m}{r} J_m  \left(k_{m,n}r\right) \hat{\phi}\right].
\end{equation}
 Analogously, 
 \begin{equation}
    \nabla \Psi^*(\vec{r},t)=\sum_{m=-\infty}^{\infty}\sum_{n=1}^{\infty} e^{i\theta_{m,n}} b_{m,n}^*(t)\frac{e^{-im\phi}}{\sqrt{\pi}R(t) J_{m+1}(z_{m,n})}  \left[\frac{\partial}{\partial r}  J_m  \left(z_{m,n}r\right) \hat{r} -  \frac{im}{r} J_m  \left(k_{m,n}r\right) \hat{\phi}\right].
\end{equation}
\begin{equation}
\begin{split}
\rho_3(\vec{r},t)&= \frac{h^2}{2\mu}\sum_{m=-\infty}^{\infty}\sum_{n=1}^{\infty} \sum_{m'=-\infty}^{\infty}\sum_{n'=1}^{\infty}  e^{(m-m')i \phi}\frac{ b_{m,n}(t)b_{m',n'}^*(t) }{\pi R(t)^2 J_{m+1 } \left(z_{m',n'}\right)J_{m'+1}  \left(z_{m',n'}\right)} e^{i({\theta_{m,n'}(t)-\theta_{m,n}(t)})} \\&\times\left[\frac{\partial}{\partial r}  J_m  \left(k_{m,n}r\right) \frac{\partial}{\partial r} J_{m'}  \left(k_{m',n'}r\right) +   \frac{mm'}{r^2} J_m  \left(k_{m,n}r\right) J_{m'}\left(k_{m',n'}r\right)  \right]
\end{split}
\end{equation}
we need to know $\rho(R_0,0)=\int \rho_3(\vec{r},t) \delta(r-R_0)r drd\phi $. We know that $\int_{0}^{2\pi} e^{(m-m')i \phi} d\phi=2\pi\delta_{m,m'}$, thus we have 
\begin{equation}
  \rho_3^E(R_0,0)=\frac{h^2}{2\mu} \sum_{m=-\infty}^{\infty}\sum_{n=1}^{\infty} \sum_{n'=1}^{\infty}  \frac{2 b_{m,n}(0)b_{m,n'}^*(0) z_{m,n}z_{m,n'}}{R_0^3  J_{m+1 } \left(z_{m,n'}\right)J_{m+1}  \left(z_{m,n'}\right)}\left[ J'_m  \left(z_{m,n}\right)  J'_{m}  \left(z_{m,n'}\right) \right]. 
\end{equation}
Applying the same recurrence relation as above we get, 
\begin{equation}
   \rho_3^E(R_0,0)= \frac{h^2}{\mu R_0^3 } \sum_{m=\infty}^{\infty}  \sum_{n=1}^{\infty} \sum_{n'=1}^{\infty} b_{m,n}(0)b_{m,n'}^*(0) z_{m,n}z_{m,n'}. 
\end{equation}
\section{Case for a spherical moving wall}
Consider a particle  with mass $\mu $ in an infinite spherical potential moving well 
\begin{equation}
    \begin{array}{cc}
 V(r)= &  \left\{
    \begin{array}{cc}
      0  \quad\text{if}& r\leq  R(t) \\
      \infty  \quad\text{if}& r>R(t)
    \end{array}
    \right.
\end{array}
\end{equation}
where $R(t)=R_0+vt$. The Hamiltonian operator is 
\begin{equation}
    \mathcal{H}=-\frac{\hbar^2}{2\mu } \left[\frac{1}{r^2} \frac{\partial }{\partial r} r^2 \frac{\partial }{\partial r} -\frac{1}{\hbar ^2r^2} \hat{L}^2\right], 
\end{equation}
where $\hat{L}$ is the orbital angular momentum operator, which is given by
\begin{equation}
 \hat{L}^2 =-\hbar^2\left[\frac{1}{\sin(\theta)}\frac{\partial }{\partial \theta} \sin(\theta) \frac{\partial }{\partial \theta} +\frac{1}{\sin^2(\theta)} \frac{\partial ^2}{\partial \phi^2}  \right] ,  
\end{equation}
in this case, the instant eigenfunctions are 
\begin{equation}
    \phi_{n,l,m}(r,\theta,\psi,t)=\frac{\sqrt{2}}{R(t)^{3/2} j_{l+1}(z_{l,n})}j_l\left(k_{l,n} r \right) Y_{l,m}(\theta, \phi),  
 \qquad k_{l,n}=\frac{z_{l,n}}{R(t)}, 
\end{equation}

where $z_{l,n}$ is the nth zero of spherical Bessel function of order $l$ ($j_l(x)$) and $Y_{l,m}(\theta, \phi) $ are the spherical harmonics. The wave function takes the form of

\begin{equation}
    \Psi(\vec{r},t)= \sum_{l=0}^{\infty} \sum_{m=-l}^{l}\sum_{n=1}^{\infty}b_{n,l,m}(t) e^{-i\theta_{l,n} (t)} \frac{\sqrt{2}}{R(t)^{3/2} j_{l+1}(z_{l,n})}j_l\left(k_{l,n} r \right) Y_{l,m}(\theta, \phi),
\end{equation}
where $\theta_{l,n} (t)= \frac{1}{\hbar} \int_{0}^{t} E_{l,n} (t')dt'$, with 
\begin{equation}
    E_{l,n}(t)= \frac{\hbar ^2 z_{l,n}^2}{ 2\mu R(t)^2}
\end{equation}
Then, the expected energy is given by 
\begin{equation*}
\langle\Psi|\hat{H}|\Psi\rangle\equiv\mathcal{E}(t)=\sum_{l=0}^{\infty} \sum_{m=-l}^{l}\sum_{n=1}^{\infty}|b_{n,l,m}(t)|^2E_{l,n}(t) , 
\end{equation*}
and its derivative in $t=0$
\begin{equation}
 \dot{\mathcal{E}}(0)=\sum_{l=0}^{\infty}  \sum_{m=l}^{m=-l} \sum_{n=1}^{\infty}\Big\{E_{l,n}(0) \Big(b^*_{n,l,m}(0)\dot{b}_{n,l,m}(0)+\dot{b}^*_{n,l,m}(0)b_{n,l,m}(0) \Big) +\dot{E}_{l,n}(0)|b_{n,l,m}(0)|^2 \Big\},
\end{equation}
where $\dot{b}(t)$ and $\dot{E}_{l,n}(t)$ are  respectively
\begin{equation}
    \dot{E}_{l,n}=-v\frac{\hbar ^2 z_{l,n}^2}{ \mu  R(t)^3}, 
\end{equation}
\begin{equation}
    \dot{b}_{n,l,m}(t)=-\sum_{l'=0}^{\infty} \sum_{m'=-l}^{l} \sum_{n'=1}^{\infty} b_{n',l',m' }(t) e^{i({\theta_{l,n}(t)-\theta_{l',n'}(t)})} \int_{0}^{R(t)} \int_{0}^{\pi} \int_{0}^{2 \pi }  \phi_{n,l,m}(r,\theta,\psi,t)\dot{\phi}_{n',l',m'}(r,\theta,\psi,t) r^2drd\Omega \label{bdot2},
\end{equation}
where $d\Omega$ is the solid angle differential. Analogously to the 2-dimensional case,  we only work when  $n'\neq n$  because when $n=n'$ the above integral vanishes. Now,
\begin{equation}
   \int_{0}^{R(t)} \int_{0}^{\pi} \int_{0}^{2 \pi }  \phi_{n,l,m}(r,\theta,\psi,t)\dot{\phi}_{n',l',m'}(r,\theta,\psi,t) r^2drd\Omega = \delta_{l,l'}\delta_{m,m'}  \frac{2}{R(t)^3 j_{l+1}(z_{l,n}) j_{l'+1}(z_{l',n'})}\int_{0}^{R(t)} j_l\left(k_{l,n} r \right) \dot{j}_{l'}\left( k_{l',n'} r\right) r^2dr, 
\end{equation}
thus we have 
\begin{equation}
\begin{split}
\sum_{l=0}^{\infty} \sum_{m=-l}^{l} \sum_{n=1}^{\infty}\ E_{l,n}(t) \dot{b}_{n,l,m}(t)b^*_{n,l,m}(t)&=  - \sum_{l=0}^{\infty} \sum_{m=-l}^{l} \sum_{ n'=1}^{\infty}\sum_{ n=1}^{\infty}b_{l,n',m}(t)b^*_{n,l,m}(t)E_{l,n}(t)  e^{i({\theta_{l,n}(t)-\theta_{l,n'}(t)})} \times\\ & \qquad\frac{2}{R(t)^3 j_{l+1}(z_{l,n}) j_{l'+1}(z_{l,n'})} \int_{0}^{R(t)} j_l\left(k_{l,n} r \right) \dot{j}_{l}\left( k_{l,n'} r\right) r^2dr, 
\\ & =-\sum_{l=0}^{\infty} \sum_{m=-l}^{l} \sum_{ n'=1}^{\infty}\sum_{ n=1}^{\infty}b^*_{n',l,m}(t)b_{n,l,m}(t) E_{l, n'}(t)  e^{i({\theta_{l,n'}(t)-\theta_{l,n}(t)})} \times \\ & \qquad \frac{2}{R(t)^3 j_{l+1}(z_{l,n}) j_{l'+1}(z_{l',n'})} \int_{0}^{R(t)} j_l\left(k_{l,n'} r \right) \dot{j}_{l}\left( k_{l,n} r\right) r^2dr,
\end{split}
\end{equation}
and
\begin{equation}
\begin{split}
     \sum_{l=0}^{\infty} \sum_{m=-l}^{l} \sum_{n=1}^{\infty} E_{l,n}(t) \dot{b}^*_{n,l,m}(t)b_{n,l,m}(t)& = - \sum_{l=0}^{\infty} \sum_{m=-l}^{l} \sum_{ n'=1}^{\infty}\sum_{ n=1}^{\infty} b_{n,l,m}(t)b^*_{n',l,m}(t)E_{l,n}(t)  e^{i({\theta_{l,n'}(t)-\theta_{l,n}(t)})} \times \\ & \qquad \frac{2}{R(t)^3 j_{l+1}(z_{l,n}) j_{l'+1}(z_{l,n'})} \int_{0}^{R(t)} j_l\left(k_{l,n} r \right) \dot{j}_{l}\left( k_{l,n'} r\right) r^2dr.
\end{split}
\end{equation}

Then, 
\begin{equation}
   \frac{2}{R(t)^3 j_l(z_{l,n}) j_l'(z_{l',n'})} \int_{0}^{R(t)} j_l\left(k_{l,n} r\right) \dot{j}_l\left(k_{l,n'} r\right) r^2dr=  -\frac{2 z_{l,n'}  v}{J_{l+3/2} \left(z_{l,n}\right)  J_{l+3/2} \left(z_{l,n'}\right)R(t)^4}\int_{0}^{R(t)} J_{l+1/2} \left(k_{l,n} r \right)  J'_{l+1/2} \left
    (k_{l,n'} r\right) r^2 dr
\end{equation}
using the same relations as above for this type of integrals we have, 
 \begin{equation}
     \int_{0}^{R(t)} J_{l+1/2} \left(k_{l,n} r\right)  J'_{l+1/2} \left(k_{l,n' } r\right) r^2dr= -\frac{R^3(t) z_{l,n}}{z_{l,n'}^2-z_{l,n}^2} J_{l+3/2} \left(z_{l,n}\right)  J'_{l+1/2} \left(z_{l,n'}\right). 
     \label{eqn:jj}
 \end{equation}
By the recurrence relations 
\begin{equation}
  \int_{0}^{R(t)} J_{l+1/2} \left(k_{l,n} r\right) J_{l+1/2}'  \left(k_{l,n' } r\right) r^2dr= \frac{R^3(t) z_{l,n}}{z_{l,n'}^2-z_{l,n}^2}  J'_{l+1/2} \left(z_{l,n}\right)  J'_{l+1/2} \left(z_{l,n'}\right),
 \end{equation}
hence, 
\begin{equation}
  \int_{0}^{R(t)} J_{l+1/2} \left(k_{l,n'} r\right) J_{l+1/2}'  \left(k_{l,n } r\right) r^2dr=-\frac{R^3(t) z_{l,n}}{z_{l,n}^2-z_{l,n'}^2}  J'_{l+1/2} \left(z_{l,n}\right)  J'_{l+1/2} \left(z_{l,n'}\right),
 \end{equation}
then  $\dot{\mathcal{E}}(0)$ is
\begin{equation}
      \dot{\mathcal{E}}(0)=-\frac{\hbar^2}{\mu R_0^3} v \left\{\sum_{l=0}^{\infty}\sum^{m=l}_{m=-l}\sum_{n=1}^{\infty}z_{l,n}^2|b_{n,l,m}(0)|^2)+ \sum_{l=0}^{\infty} \sum^{m=l}_{m=-l}\sum_{n=1}^{\infty}\sum_{n'\neq n}^{\infty} b^*_{n,l,m}(0)b_{n',l,m}(0) z_{l,n}z_{l,n'} \frac{J'_{l+1/2} \left(z_{l,n}\right)  J'_{l+1/2} \left(z_{l,n'}\right)}{J_{l+3/2} \left(z_{l,n}\right)  J_{l+3/2} \left(z_{l,n'}\right)} \right\}, 
\end{equation}
that can be written as
\begin{equation}
      \dot{\mathcal{E}}(0)=-\frac{\hbar^2}{\mu R_0^3} v \left\{\sum_{l=0}^{\infty}\sum_{m=-l}^{m=l}\sum_{n=1}^{\infty}z_{l,n}^2|b_{n,l,m}(0)|^2+ \sum_{l=0}^{\infty} \sum_{m=-l}^{m=l} \sum_{n=1}^{\infty}\sum_{n'\neq n}^{\infty}b^*_{n,l,m}(0)b_{n',l,m}(0) z_{l,n}z_{l,n'} \right\}. 
\end{equation}
The energy density in this case is
\begin{equation}
\rho_3(\vec{r},t)= \frac{\hbar^2}{2\mu } \nabla \Psi^*(\vec{r},t)\cdot \nabla \Psi(\vec{r},t) 
\end{equation}
where
\begin{equation}
    \nabla \Psi(\vec{r},t)=\sum_{l=0}^{\infty}\sum_{m=-l}^{m=l} \sum_{n=1}^{\infty}e^{-i\theta_{l,n}} b_{n,l,m}(t) \frac{\sqrt{2}}{R(t)^{3/2} j_{l+1}(k_{l,n}a)}  \left[\frac{\partial}{\partial r}  j_l  \left(k_{l,n}r\right) \mathbf{Y}_{l,m} +   \frac{1}{r} j_l  \left(k_{l,n}r\right) \mathbf{\Psi}_{l,m}\right]
\end{equation}
 where $ \mathbf{Y}_{l,m}= Y_{l,m}(\theta,\psi)\hat{r} $ and $\mathbf{\Psi}_{l,m}=r \nabla Y_{l,m}(\theta,\psi)$ are the vector spherical harmonics\cite{RBarrera_1985}, this follow the orthogonality relation $\mathbf{\Psi}_{l,m}\cdot\mathbf{Y}_{l,m} $ =0. Analogously, 
 \begin{equation}
    \nabla \Psi^*(\vec{r},t)=\sum_{l=0}^{\infty}\sum_{m=-l}^{m=l} \sum_{n=1}^{\infty}e^{i\theta_{l,n}} b_{n,l,m}^*(t) \frac{\sqrt{2}}{R(t)^{3/2} j_{l+1}(k_{l,n}a)}  \left[\frac{\partial}{\partial r}  j_l  \left(k_{l,n}r\right) \mathbf{Y}_{l,m}^* +   \frac{1}{r} j_l  \left(k_{l,n}r\right) \mathbf{\Psi}_{l,m}^*\right]
\end{equation}
thus, 
\begin{equation}
\begin{split}
\rho_3(\vec{r},t)&= \frac{h^2}{2\mu }\sum_{l=0}^{\infty}\sum_{m=-l}^{m=l}\sum_{n=1}^{\infty}\sum_{l'=0}^{\infty}\sum_{m'=-l'}^{m'=l'}  \sum_{n'=1}^{\infty}\frac{2 b_{n,l,m}(t)b_{n',l',m'}^*(t) }{R(t)^3  j_{l+1 } \left(z_{l',n'}\right)j_{l'+1}  \left(z_{l',n'}\right)} e^{i({\theta_{l,n'}(t)-\theta_{l,n}(t)})} \\&\left[\frac{\partial}{\partial r}  j_l  \left(k_{l,n}r\right) \frac{\partial}{\partial r} j_{l'}  \left(k_{l',n'}r\right) \mathbf{Y}_{l',m'}^* \cdot \mathbf{Y}_{l,m}+   \frac{1}{r^2} j_l  \left(k_{l,n}r\right) j_{l'}\left(k_{l',n'}r\right) \mathbf{\Psi}_{l,m}^*\cdot \mathbf{\Psi}_{l',m'} \right]
\end{split}
\end{equation}
we want to know the total force exerted by the particle on the boundary $\mathcal{F}=\int_{0}^{2\pi}\int_{0}^{\pi} \rho_3(\vec{r},t) \delta(r-R_0)r^2 drd\Omega $. The vector spherical harmonics satisfy the integral equations relations $\int_{0}^{2\pi}\int_{0}^{\pi} \mathbf{\Psi}_{l,m}^*\cdot \mathbf{\Psi}_{l',m'} d\Omega=l(l+1) \delta_{l,l'}\delta_{m,m'}$ and $\int_{0}^{2\pi}\int_{0}^{\pi} \mathbf{Y}_{l,m}^*\cdot \mathbf{Y}_{l',m'}d\Omega=\delta_{l,l'}\delta_{m,m'}$, thus we have 
\begin{equation}
  \rho_3^E(R_0,0)=\frac{h^2}{2\mu }\sum_{l=0}^{\infty}\sum_{m=-l}^{m=l}   \sum_{n=1}^{\infty} \sum_{n'=1}^{\infty}\frac{2 b_{n,l,m}(0)b_{n',l,m}^*(0) z_{l,n}z_{l,n'}}{R_0^3  j_{l+1 } \left(z_{l,n'}\right)j_{l+1}  \left(z_{l,n'}\right)}\left[ j'_l  \left(z_{l,n}\right)  j'_l  \left(z_{l,n'}\right) \right]. 
\end{equation}
Finally, using the properties of the Bessel functions, we get
\begin{equation}
   \rho_3^E(R_0,0)= \frac{h^2}{\mu R_0^3 }\sum_{l=0}^{\infty}\sum_{m=-l}^{m=l} \sum_{n=1}^{\infty} \sum_{n'=1}^{\infty}  b_{n,l,m}(0)b_{n',l,m}^*(0) z_{l,n}z_{l, n'}
\end{equation}

%%%%%%%%%%%%%%%%%%%%%%%%%%%%%%%%%%%%%%%%%
\bibliographystyle{unsrt} % We choose the "plain" reference style
\bibliography{referencesTOL2023.bib} 
\end{document}